\documentclass{article}
\usepackage{amssymb}

\begin{document}

\title{Magnetic properties of weakly doped antiferromagnets}
\author{I. R. Pimentel,$^{*}$ F. Carvalho Dias and  L. M. Martelo
\\{\small Department of Physics and  CFMC, University of
  Lisbon, 1649 Lisboa, Portugal}\\
R. Orbach 
\\ {\small Department of Physics,
  University of California, Riverside, CA92521, USA}}
\date{}
\maketitle

\begin{abstract}
We study the spin excitations and the transverse susceptibility of a
two-dimensional antiferromagnet doped with a small concentration of holes in
the t-J model. The motion of holes generates a renormalization of the
magnetic properties. The Green\'{}s functions are calculated in the
self-consistent Born approximation. It is shown that the long-wavelength
spin waves are significantly softened and the shorter-wavelength spin waves
become strongly damped as the doping increases. The spin wave velocity is 
reduced by the coherent motion of holes, and not increased
as has been claimed elsewhere. The transverse susceptibility is found to
increase considerably with doping, also as a result of coherent hole motion.
Our results are in agreement with experimental data for the doped
copper oxide superconductors.
\end{abstract}

\section{Introduction}

After the discovery of high-T$_{c}$ superconductors there has been intensive
investigation of the magnetic properties$^{1}$ of doped copper oxide
materials because of their connection to high temperature superconductivity.
The undoped compounds are antiferromagnetic (AF) insulators. Doping
introduces holes,$^{2,3}$ the charge carriers, in the AF square lattice of
the copper oxide planes. The long-range AF order rapidly disappears at low
doping, and superconductivity arises upon further doping. Strong
two-dimensional AF fluctuations are nevertheless observed$^{4}$ even at
fairly high doping, suggesting a conducting phase that, in spite of being
paramagnetic, exhibits short-range AF order. A striking feature of the
copper oxides is the strong sensitivity of their magnetic properties to the
hole concentration, $\delta$. Experiments have shown important softening and
damping in the spin excitations,$^{5-7}$ as well as a significant increase
in the spin susceptibility,$^{8-10}$ for the doped copper oxides. It is
therefore important to study the interplay between doping and
antiferromagnetism for an understanding of these materials.

It is believed that the essential physics of strong electron correlations in
the copper oxide planes is described by the t-J model

\begin{equation}
H_{t-J}=-t\sum_{<i,j>,\sigma }\left( c_{i\sigma }^{\dagger }c_{j\sigma}
+H.c.\right) +J\sum_{<i,j>}\left( \mathbf{S}_{i}\cdot \mathbf{S}_{j}-
\frac{1}{4}n_{i}n_{j}\right),  \label{1}
\end{equation}
where $c_{i\sigma}^{\dagger}$ and $c_{i\sigma}$ are creation and annihilation
 electron operators acting on a reduced Hilbert space with no doubly occupied
  sites, the spin operator is $S_{i}^{\mu }=\frac{1}{2}\sum_{\alpha \beta }
c_{i\alpha }^{\dagger }\sigma_{\alpha \beta }^{\mu }c_{i\beta }$,
$n_{i}=n_{i\uparrow }+n_{i\downarrow} $ and $n_{i\sigma }=
c_{i\sigma }^{\dagger }c_{i\sigma }$. In the copper oxides, $J\simeq 1500K$
 and $t\sim 3J$. For the undoped materials, i.e. at half-filling, only the
  Heisenberg part of the Hamiltonian is relevant and it
describes a spin-1/2 AF insulator. With doping, and nearly half-filling, the
Hamiltonian describes holes moving in an AF background, the holes strongly
interacting with the spin array. The motion of holes tends to disrupt the AF
order, because a moving hole leaves behind a string of flipped spins. The
charge carriers are then holes dressed by a cloud of spin excitations.

The propagation of a single hole in a two dimensional antiferromagnet has
been studied with a variety of approaches. Considering the t-J model in a
Schwinger boson representation, hole motion was treated within a
self-consistent Born approximation (SCBA).$^{11-15}$ It was found that a
hole can propagate coherently because of its strong coupling to the spin
excitations, having a quasi-particle bandwidth $\thicksim J$ and energy
minima at momenta $\mathbf{q}_{i}=\left( \pm \pi /2,\pm \pi /2\right) $. The
calculated spectral density shows a quasi-particle peak of intensity
$\thicksim (J/t)^{2/3}$ and a broad incoherent multiple spin wave continuum,
of width $\thicksim 2zt$ ($z$ is the coordination number), that
is located at higher energies. These results are in good agreement with
those from exact diagonalization of small clusters.$^{16}$ The study of hole
motion has been extended for a finite concentration of holes, within the
SCBA.$^{17-19}$ The results obtained show that, to first order in $\delta$, the
quasiparticle characteristics remain essentially the same, supporting a
description of the quasi-holes as noninteracting particles, filling up a
Fermi surface consisting of four pockets located at momenta $\mathbf{q}_{i}$,
 having an enclosed area proportional to $\delta$. The spectral density for
finite doping again contains a coherent and an incoherent part. The
role of the coherent motion of holes on the magnetic properties of
the copper oxides has been a matter of discussion. Some authors,$^{17,20}$
 in contradiction to others,$^{21-23}$ have claimed that the coherent 
motion of holes leads to stiffening of the spin excitations, while it is 
the incoherent motion of holes that generates significant softening, leading
 to an overall softening.

In this work we study the effects of hole doping on the spin
excitations, calculating both the softening and the damping, and 
determine the doping dependence of the transverse spin susceptibility of a
two-dimensional antiferromagnet, discussing in particular the contributions
from the coherent and the incoherent motion of holes. Our study starts from
the t-J model in the Schwinger boson representation and is carried out in
the SCBA. It is shown that the magnetic properties are very sensitive to
hole doping, as a result of the strong interaction between the hole and spin
systems, and that the coherent motion of holes leads in fact to softening of
 the spin excitations.

\section{ The Interaction Between Holes And Spin Waves}

Our system is described by the t-J Hamiltonian $(1)$ on a two dimensional
square lattice. In order to enforce no double occupancy of sites we use the
slave-fermion Schwinger Boson representation $c_{i\sigma }=f_{i}^{\dagger}
b_{i\sigma }$, where the slave-fermion operator $f_{i}^{\dagger }$ creates
a hole and the boson operator $b_{i\sigma }$ accounts for the spin, subject
to the constraint $f_{i}^{\dagger}f_{i}+\sum_{\sigma }b_{i\sigma}^{\dagger}
b_{i\sigma }=2S$.

We consider the low doping regime, $\delta\ll 1$, where the states are close to
the pure AF state, and hence exhibit long-range order. The AF state is
approximated by the N\'{e}el state, which in the Schwinger representation
can be interpreted as a condensate of Bose fields $b_{i\uparrow }=\sqrt{2S}$
and $b_{j\downarrow }=\sqrt{2S}$, respectively in the up and down
sub-lattices, and the bosons $b_{i\downarrow }=b_{i}$ and $b_{j\uparrow}=b_{j}$
 are then Holstein-Primakov spin-wave operators on the N\'{e}el state.

The Hamiltonian $(1)$, with $S=1/2$, then becomes

\begin{eqnarray}
H_{t-J} &=&-t\sum_{<i,j>}\left[ f_{i}f_{j}^{\dagger }\left( b_{i}^{\dagger
}+b_{j}\right) +H.c.\right]  \nonumber \\
&&+\frac{J}{2}\sum_{<i,j>}\left( 1-f_{i}^{\dagger }f_{i}\right) \left(
1-f_{j}^{\dagger }f_{j}\right) \left[ b_{i}^{\dagger }b_{i}+b_{j}^{\dagger
}b_{j}+b_{i}b_{j}+b_{i}^{\dagger }b_{j}^{\dagger }-\frac{1}{2}\right].\hspace{0.5cm}
\label{(2)}
\end{eqnarray}
The transfer part describes the reversal of spins as the hole
moves. In the Heisenberg part, the factor $\left( 1-f_{i}^{\dagger}f_{i}\right)
 \left( 1-f_{j}^{\dagger }f_{j}\right)$ accounts for a loss of
magnetic energy due to doping, and for small hole concentrations it may be
replaced by $1-f_{i}^{\dagger }f_{i}-f_{j}^{\dagger }f_{j}$.

Applying Fourier transforms and the Bogoliubov-Valatin transformation for
the spin variables: $b_{-\mathbf{k}}^{\dagger }=u_{\mathbf{k}}
\beta _{-\mathbf{k}}^{\dagger }+v_{\mathbf{k}}\beta _{\mathbf{k}}$ and
 $b_{\mathbf{k}}=v_{\mathbf{k}}\beta _{-\mathbf{k}}^{\dagger }+ u_{\mathbf{k}}
 \beta _{\mathbf{k}}$, where $u_{\mathbf{k}}= \left[ \left( (1-
 \gamma _{\mathbf{k}}^{2})^{-1/2}+1\right) /2\right] ^{1/2}$,
  $v_{\mathbf{k}}=-\mathrm{sgn}(\gamma_{\mathbf{k}})\left[ \left( (1-
 \gamma _{\mathbf{k}}^{2})^{-1/2}-1\right)
/2\right] ^{1/2}$, and $\gamma _{\mathbf{k}}=\frac{1}{2}\left( \cos
k_{x}+\cos k_{y}\right) $, we obtain from $(2)$ the effective Hamiltonian

\begin{equation}
H=-\frac{1}{\sqrt{N}}\sum_{\mathbf{q},\mathbf{k}}f_{\mathbf{q}}f_{\mathbf{q}-
\mathbf{k}}^{\dagger }\left[ V(\mathbf{q},-\mathbf{k})\beta _{-\mathbf{k}}+
V(\mathbf{q-k},\mathbf{k})\beta _{\mathbf{k}}^{\dagger }\right]
+\sum_{\mathbf{k}}\omega _{\mathbf{k}}^{0}\beta _{\mathbf{k}}^{\dagger }
\beta _{\mathbf{k}}
.\label{3}
\end{equation}
Here, $V(\mathbf{q},\mathbf{k})=zt\left( \gamma _{\mathbf{q}}u_{\mathbf{k}}
+\gamma _{\mathbf{q}+\mathbf{k}}v_{\mathbf{k}}\right) $, $\omega
_{\mathbf{k}}^{0}=(zJ/2)\left( 1-\gamma _{\mathbf{k}}^{2}\right) ^{1/2}$, the
coordination number is $z=4$, the sums run over the Brillouin zone for an
antiferromagnet on a square lattice, and $N$ is the number of sites in each
sub-lattice. In $(3)$, the first term represents the interaction between
holes and spin waves resulting from the motion of holes with emission and
absorption of spin waves, and the second term describes spin waves in a pure
antiferromagnet. In writing $(3)$ we neglected an interaction term involving
the scattering of holes by spin-waves, proportional to $J$, because its
effect is small compared to the other term, proportional to $t$.$^{17}$ We
note that at the bare level the holes have no dispersion. In fact, they
propagate only after being dressed by spin waves. Here we study the
renormalization of the magnetic properties induced by the dynamical
interaction between the holes and the spin waves.

\section{Green's Functions For Spin Waves And Holes}

Given the magnitude of the couplings in the copper oxides, we make
use of the Green's function formalism at zero-temperature. The Green's
functions for the spin waves are defined as

\[
\begin{array}{l}
D^{-+}(\mathbf{k},t-t^{\prime })=-i\left\langle \mathcal{T}\beta _{\mathbf{k}
}(t)\beta _{\mathbf{k}}^{\dagger}(t^{\prime })\right\rangle, \\ 
D^{+-}(\mathbf{k},t-t^{\prime })=-i\left\langle \mathcal{T}
\beta _{-\mathbf{k}}^{\dagger}(t)\beta _{-\mathbf{k}}(t^{\prime })
\right\rangle, \\
D^{--}(\mathbf{k},t-t^{\prime })=-i\left\langle \mathcal{T}\beta _{\mathbf{k}
}(t)\beta _{-\mathbf{k}}(t^{\prime })\right\rangle, \\ 
D^{++}(\mathbf{k},t-t^{\prime })=-i\left\langle \mathcal{T}\beta _{-\mathbf{k
}}^{\dagger}(t)\beta _{\mathbf{k}}^{\dagger}(t^{\prime })\right\rangle,
\end{array}
\]
where $\left\langle \quad \right\rangle $ represents an average
over the ground state. Their Fourier transforms satisfy the Dyson equations:

\[
D^{\mu \nu }(\mathbf{k},\omega )=D_{0}^{\mu \nu }(\mathbf{k},\omega
)+\sum_{\gamma \delta }D_{0}^{\mu \gamma }(\mathbf{k},\omega )\Pi ^{\gamma
\delta }(\mathbf{k},\omega )D^{\delta \nu }(\mathbf{k},\omega ),
\]
where $\mu ,\nu =\pm $. The free Green's functions are
\[
\begin{array}{l}
D_{0}^{-+}(\mathbf{k},\omega )=\left( \omega -\omega _{\mathbf{k}}^{0}+i\eta
\right) ^{-1}\qquad, \\ 
D_{0}^{+-}(\mathbf{k},\omega )=\left( -\omega -\omega _{\mathbf{k}}^{0}+
i\eta \right) ^{-1}, \\
D_{0}^{--}(\mathbf{k},\omega )=D_{0}^{++}(\mathbf{k},\omega )=0\qquad,
\end{array}
\]
with $\eta \rightarrow 0^{+}$, and $\Pi ^{\gamma \delta }(\mathbf{k},\omega )$
are the self-energies generated by the interaction between holes
and spin waves.

We calculate the self-energies in the SCBA, corresponding to only ''bubble''
diagrams with dressed hole propagators. These diagrams describe the decay of
spin-waves into ''particle-hole'' pairs. The approximation neglects
corrections to the hole-spin interaction vertex, which have been shown to be
unimportant.$^{12,15,17}$ The self-energies are given by

\begin{equation}
\Pi ^{\gamma \delta }(\mathbf{k},\omega )=-i\frac{1}{N}\sum_{\mathbf{q}}
U^{\gamma \delta }(\mathbf{k},\mathbf{q})\int_{-\infty }^{+\infty }
\frac{d\omega _{\mathbf{q}}}{2\pi }G(\mathbf{q},\omega _{\mathbf{q}})
G(\mathbf{q}-\mathbf{k},\omega _{\mathbf{q}}-\omega ),  \label{4}
\end{equation}
where $G(\mathbf{q},\omega _{\mathbf{q}})$ is the Fourier
transform of the Green's function for the dressed holes, $G(\mathbf{q},t-
t^{\prime})=
-i\left\langle \mathcal{T}f_{\mathbf{q}}(t)f_{\mathbf{q}}^{\dagger}(t^{\prime})
\right\rangle $, and

\begin{eqnarray*}
U^{+-}(\mathbf{k},\mathbf{q}) &=&V(\mathbf{q}-\mathbf{k},\mathbf{k})^{2}\quad ,
\quad U^{-+}(\mathbf{k},\mathbf{q})=V(\mathbf{q},-\mathbf{k})^{2},
\\
U^{--}(\mathbf{k},\mathbf{q}) &=&U^{++}(\mathbf{k},\mathbf{q})=V(\mathbf{q},
-\mathbf{k})V(\mathbf{q}-\mathbf{k},\mathbf{k}).
\end{eqnarray*}
The relations $\Pi ^{-+}(\mathbf{k},\omega )=\Pi ^{+-}(-\mathbf{k},-\omega )$
and $\Pi ^{--}(\mathbf{k},\omega )=\Pi ^{++}(\mathbf{k},\omega )$
are verified.

In the SCBA the holes are dressed by pure AF spin waves. This approach
implies a spectral function for the holes that is composed of a coherent
quasi-particle peak and an incoherent continuum, with the quasi-holes
filling up a Fermi surface that consists of pockets located at $\mathbf{q}_{i}=
(\pm \pi /2,\pm \pi /2)$, as mentioned above. We shall
take for the hole spectral function the approximate form
 
\begin{equation}
\rho (\mathbf{q},\omega )=\left[ \rho^{coh}(\mathbf{q},\omega )+\rho^{incoh}
(\mathbf{q},\omega )\right]
\mathcal{F}^{\pm}(\mathbf{q})\theta (\pm \omega),  \label{5}
\end{equation}
with, Fermi surface $\mathcal{F}^{-}(\mathbf{q})=\sum_{i=1}^{4}\theta(q_{F}-|\mathbf{q}-\mathbf{q}_{i}|)$, $\mathcal{F}^{+}(\mathbf{q})=1-\mathcal{F}^{-}(\mathbf{q})$, Fermi momentum  $q_{F}=\sqrt{\pi \delta}$, and

\begin{eqnarray*}
\rho^{coh}(\mathbf{q},\omega ) &=&a_{0}\delta (\omega -\varepsilon
_{\mathbf{q}}), \\
\rho^{incoh}(\mathbf{q},\omega ) &=&h\theta (|\omega |-zJ/2)\theta
(2zt+zJ/2-|\omega |).
\end{eqnarray*}
Here the energies are measured with respect to the Fermi level,
and the quasi-particle dispersion can, near the minima at $\mathbf{q}_{i}$,
be written as $\varepsilon _{\mathbf{q}}=\varepsilon_{min}+
(\mathbf{q}-\mathbf{q}_{i})^{2}/2m$, with an effective mass $m\simeq 1/J$
(neglecting band anisotropy). The quasi-particle residue is $a_{0}\simeq
(J/t)^{2/3}$,
 and the remaining spectral density appears in the
incoherent continuum of width $2zt$ and height $h\simeq\left(1-a_{0}\right)
/2zt$, satisfying the sum rule $\int d\omega\rho(\mathbf{q},\omega)=1$.

The spin wave self-energies $(4)$ are obtained in terms of the hole
spectral function $(5)$ by

\begin{equation}
\Pi ^{\gamma \delta }(\mathbf{k},\omega )=\frac{1}{N}\sum_{\mathbf{q}}U^{\gamma \delta }
(\mathbf{k},\mathbf{q}) \left[ Y(\mathbf{q},-\mathbf{k};\omega )+
Y(\mathbf{q}-\mathbf{k},\mathbf{k};-\omega )\right],  \label{6}
\end{equation}
with

\[
Y(\mathbf{q},-\mathbf{k};\omega )=\int_{0}^{+\infty }d\omega ^{\prime
}\int_{-\infty }^{0}d\omega ^{\prime \prime }\frac{\rho (\mathbf{q},
\omega ^{\prime })\rho (\mathbf{q}-\mathbf{k},\omega ^{\prime \prime })}
{\omega +\omega ^{\prime \prime }-\omega ^{\prime }+i\eta }.
\]
From $(5)$ and $(6)$ follows that the self-energies will present
three contributions,

\[
\Pi ^{\gamma \delta }(\mathbf{k},\omega )=\Pi _{c,c}^{\gamma \delta }
(\mathbf{k},\omega )+\Pi _{c,ic}^{\gamma \delta }(\mathbf{k},\omega )+\Pi
_{ic,ic}^{\gamma \delta }(\mathbf{k},\omega ),
\]
corresponding, respectively, to transitions of holes within the
coherent band, between the coherent and incoherent bands, and within the
incoherent band. We have calculated these different contributions to lowest
order in the hole concentration $\delta$. The imaginary parts of these
contributions are non-zero only in certain regions of the $(\mathbf{k},\omega)$
 space: $\mathrm{Im}\Pi _{c,c}\neq 0$ for $\left[
-kq_{F}/m+k^{2}/2m\right] <\omega <\left[ kq_{F}/m+k^{2}/2m\right] $,
$\mathrm{Im}\Pi_{c,ic}\neq 0$ for $zJ/2<\omega <zJ/2+2zt$, and
$\mathrm{Im}\Pi_{ic,ic}\neq 0$ for $zJ<\omega <zJ+4zt$.

\section{Magnetic Properties }

We now present the calculation of the effects of hole doping on the
magnetic properties.

The renormalized spin wave energy $\omega _{\mathbf{k}}$ is given by the
poles of the Green's functions $D(\mathbf{k},\omega )$, determined by the
condition

\[
\left[ (D_{0}^{-+})^{-1}-\Pi ^{+-}\right] \left[ (D_{0}^{+-})^{-1}-\Pi
^{-+}\right] -\Pi ^{++}\Pi ^{--}=0.
\]

In the region where $\mathrm{Im}\Pi (\mathbf{k},\omega )=0$, we find to lowest
order in $\delta$,

\begin{equation}
\;\;\omega _{\mathbf{k}} =
\omega _{\mathbf{k}}^{0}+\mathrm{Re}\Pi ^{+-}
(\mathbf{k},\omega _{\mathbf{k}}^{0}),  \label{7}
\end{equation}
leading to
\begin{equation}
\omega _{\mathbf{k}} =
\omega _{\mathbf{k}}^{0}(1-r(\mathbf{k})),  \label{8}
\end{equation}
where

\[
r(\mathbf{k}) =\delta a_{0}^{2}\left(\frac{t}{J}\right)^{2}\left\{
\frac{1}{2}\left(\frac{k^{2}}{1-\gamma_{\mathbf{k}}^{2}}\right)
\theta(2q_{F}-k)+\right.\hspace{3.cm}
\]
\[
\left.+
\left(\frac{sin^{2}k_{x}+sin^{2}k_{y}}{1-\gamma_{\mathbf{k}}^{2}}\right)
\left(\frac{1-\gamma_{\mathbf{k}}^{2}-(k/2)^{2}}{1-\gamma_{\mathbf{k}}^{2}
-(k/2)^{4}}\right)\theta(k-2q_{F})\right\}+
\]
\[
+\sqrt{\delta}\frac{t}{J}\frac{(1-a_{0})^{2}}{2}\left[ 
\ln 2 +\frac{a_{0}}{1-a_{0}}\ln\left(1+4\frac{t}{J}\right)\right]
\left\{
\frac{1}{2\sqrt{\pi}}\left(\frac{k^{3}}{1-\gamma_{\mathbf{k}}^{2}}
\right)\theta (2q_{F}-k)+\right.
\]
\[
\left.
+\sqrt{\delta}\left(\frac{sin^{2}k_{x}+sin^{2}k_{y}}
{1-\gamma_{\mathbf{k}}^{2}}\right)\theta (k-2q_{F})\right\}.
\]
In $r(\mathbf{k})$ the first term is
generated only by the coherent
motion of holes, i.e. $\Pi _{c,c}^{+-}$, whereas the second involves the
incoherent motion resulting from the sum $\Pi _{c,ic}^{+-}+\Pi _{ic,ic}^{+-}$.
One finds that both the coherent and the incoherent motion of holes generate a reduction of the spin wave energy, and hence give rise to softening of the spin excitations. The fact that the coherent motion of holes leads to softening, even in the regime where the spin wave velocity is larger than the hole Fermi velocity, is explained in detail in the Appendix.

In the long-wavelength limit, $k\ll 1$, one has

\begin{equation}
\omega _{\mathbf{k}}=ck,  \label{9}
\end{equation}
with

\begin{equation}
c=Z_{c}c_{0},  \label{10}
\end{equation}
where $c_{0}=zJ/(2\sqrt{2})$ is the spin wave velocity for a pure
antiferromagnet, and the renormalization factor is

\begin{equation}
\ Z_{c}=1-\delta a_{0}^{2}\left(\frac{t}{J}\right)^{2}.   \label{11}
\end{equation}
For finite hole concentrations one has $Z_{c}<1$, which implies
a reduction of the spin wave velocity with doping. This effect is
generated only by the coherent motion of the holes, as can be seen from $(8)$.

In the region where the spin wave dispersion crosses the pair excitation
continuum, defined as the region where $\mathrm{Im}\Pi (\mathbf{k,}\omega
)\neq 0$, one finds, to lowest order in $\delta$, that the spin excitations
become damped, acquiring an inverse lifetime given by 
\begin{equation}
\Gamma (\mathbf{k})=-2\mathrm{Im}\Pi ^{+-}(\mathbf{k},
\omega _{\mathbf{k}})  \label{(12)}
\end{equation}
One finds that the damping is determined only by the coherent motion of
holes, i.e., $\mathrm{Im}\Pi _{c,c}^{+-}$ , because the contributions 
involving the
incoherent motion, $\mathrm{Im}\Pi _{c,ic}^{+-}$ and 
$\mathrm{Im}\Pi _{ic,ic}^{+-}$,
vanish in the relevant region of the $(\mathbf{k},\omega )$ space. Hence we
 have
\begin{eqnarray}
\;\Gamma (\mathbf{k}) =zJ\sqrt{\delta} a_{0}^{2}\left(\frac{t}{J}\right)^{2}
\frac{1}{\sqrt{\pi}k
(1-\gamma_{\mathbf{k}}
^{2})^{1/2}}F^{+-}(\mathbf{k})\times \hspace{3.cm} \label{13} \\
\qquad \left[ \sqrt{1-s^{2}(g_{\mathbf{k}})}\theta (1-|s(g_{\mathbf{k}})|)
-\sqrt{1-s^{2}(-g_{\mathbf{k}})}
\theta(1-|s(-g_{\mathbf{k}})|)\right],  \nonumber
\end{eqnarray}

with

\[
F^{+-}(\mathbf{k}) = (\cos k_{y}-\cos k_{x})\left( \cos
(g_{\mathbf{k}}k_{x})-\cos (g_{\mathbf{k}}k_{y})\right)+\hspace{2.5cm}
\]
\[
-2(1-\gamma _{\mathbf{k}}^{2})^{1/2}\left( \sin k_{x}\sin
(g_{\mathbf{k}}k_{x})+\sin k_{y}\sin (g_{\mathbf{k}}k_{y})\right) 
+4(1-\gamma _{\mathbf{k}}^{2}),
\]
where $s(g_{\mathbf{k}})=(1-g_{\mathbf{k}})k/2q_{F}$,$\;g_{\mathbf{k}}=
(2\omega _{\mathbf{k}}/J$\strut $)/k^{2}$,
and $\omega _{\mathbf{k}}$ is given by $(8)$. We note the strong
doping dependence of the damping $\thicksim \sqrt{\delta}$, as compared 
to that of the reduction of the spin wave velocity $(Z_{c}-1)\thicksim \delta$, $\delta\ll 1$.

From $(13)$ one has that for sufficiently small doping, long-wavelength
spin waves remain well defined, whereas the shorter wavelength spin waves
are damped, decaying into ''particle-hole'' pairs. As the doping increases
more spin waves, in the shorter wavelength side, dive into the pair
excitation continuum and become damped. For hole concentrations above a
certain threshold $\delta^{*}$, such that the spin wave velocity equals the
Fermi velocity, $Z_{c}^{*}c_{0}/(k_{F}^{*}/m)=1$, the spin wave dispersion lies
entirely in the pair excitation continuum, and then even the long-wavelength
spin waves are damped. In the limit $k\ll 1$ one has $\Gamma \thicksim k$,
which implies that the spin waves are overdamped.

The transverse spin susceptibility is defined by

\begin{equation}
\chi _{\bot }=\chi _{\bot }(\mathbf{k}=0,\omega =0),  \label{14}
\end{equation}
where the dynamical susceptibility is given by

\[
\chi _{\bot }(\mathbf{k},\omega )=i\int_{0}^{\infty }dte^{i\omega t}<\left[
S^{x}(\mathbf{k},t),S^{x}(-\mathbf{k},0)\right] >.
\]

Writing the spin operator in terms of the electron creation and
annihilation operators, $S_{i}^{x}$\strut $=(S_{i}^{+}+S_{i}^{-})/2$ with
$S_{i}^{+}=c_{i\uparrow }^{\dagger }c_{i\downarrow }$ and
$S_{i}^{-}=c_{i\downarrow }^{\dagger }c_{i\uparrow }$, using the Schwinger
boson representation with the bose condensation associated with the N\'{e}el
state, and performing the Bogoliubov-Valatin transformation, one finds that
the susceptibility can be expressed in terms of the spin wave Green's 
functions by

\begin{equation}
\chi _{_{_{\perp }}}=-\lim_{\mathbf{k\rightarrow 0}}\left( 
\frac{1-\gamma _{\mathbf{k}}}{1+\gamma _{\mathbf{k}}}\right) ^{1/2}\left[ 
\mathrm{Re}D^{+-}(\mathbf{k},0)+\mathrm{Re}D^{++}(\mathbf{k},0)\right].
\label{15}
\end{equation}
In $(15)$ we have approximated $<f_{i}^{\dagger
}f_{i}b_{i}^{\dagger }>\simeq \delta<b_{i}^{\dagger }>$, and neglected a
prefactor $(1-\delta)^{2}$ which is caused by dilution of the spin lattice by
holes. To lowest order in $\delta$, the transverse susceptibility is given by 

\begin{equation}
\chi _{_{_{\perp }}}=\lim_{\mathbf{k}\rightarrow 0}\frac{1}{zJ(1+
\gamma _{\mathbf{k}})}\left[ 1-\frac{2}{zJ(1-\gamma _{\mathbf{k}}^{2})^{1/2}}
\left( \mathrm{Re}\Pi ^{+-}(\mathbf{k},0)+\mathrm{Re}\Pi ^{++}(\mathbf{k},0)
\right) \right].  \label{16}
\end{equation}
One finds that only the coherent motion, i.e., $\Pi _{c,c}^{+\pm}$,
contributes to the susceptibility in $(16)$, because the contributions
involving the incoherent motion, $\Pi _{c,i}^{+\pm}$ and 
$\Pi_{i,i}^{+\pm}$, vanish in the limit $\mathbf{k}\rightarrow 0$. We
then obtain

\begin{equation}
\chi _{_{_{\perp }}}=Z_{\chi}\chi _{_{_{\perp }}}^{0},  \label{17}
\end{equation}
where $\chi _{_{_{\perp }}}^{0}=1/(2zJ)$ is the transverse
susceptibility for a pure Heisenberg antiferromagnet, and the
renormalization factor is given by

\begin{equation}
Z_{\chi}=1+4\delta a_{0}^{2}\left(\frac{t}{J}\right)^{2}.  \label{18}
\end{equation}
From $(18)$ one sees that the transverse susceptibility increases
with hole doping, this effect being determined by the coherent motion of
holes.

\section{Results and Discussion}

We have considered a two-dimensional antiferromagnet doped with a small
concentration of mobile holes and calculated the renormalization of magnetic
properties induced by hole motion. In our calculation it is assumed that
there is long-range AF order in the system. In real materials true
long-range AF order disappears at rather low concentrations, e.g.,
$\delta_{c}\simeq 0.02$ for $La_{2-\delta}Sr_{\delta}CuO_{4}$. However,
experiments have revealed that above such concentrations, there are large AF
correlated regions in the system corresponding to the size of the magnetic
correlation length $\xi $, scaling like $\xi\sim 1/\sqrt{\delta}$.$^{24}$ 
Those regions can in particular sustain spin
excitations with wavelengths up to the region size. One expects that the
results that we derived when there is long-range order, still describe the
physics on length scales less than $\xi $, with $\xi $ large, when long range
order is broken.

We find that the spin excitations are very sensitive to doping, with
significant softening and damping occurring as a result of hole 
motion. In the low momenta region, the reduction of the spin wave energy
is mainly determined by the coherent motion of holes, while the contribution
from the incoherent motion becomes more significant with increasing momenta.
The spin wave velocity decreases due to the coherent motion of holes, which for
$t/J=3$  and a concentration $\delta=0.02$ produces 
a renormalization  factor $Z_{c}=0.96$, while a concentration $\delta=0.05$ 
leads to $Z_{c}=0.90$. However, for momenta $k$ around $q_{F}$, the slope of
the spin dispersion shows a much higher reduction, by a factor $0.91$ for a
 concentration $\delta\simeq 0.02$, and a factor $0.78$ for a concentration
$\delta\simeq 0.05$, as a result of the coherent plus the incoherent 
motion of holes.
For $\delta\simeq 0.02$ there is little damping since only spin
 waves near
the upper end of the spin wave spectrum lie inside the pair excitation
continuum. For $\delta\simeq 0.05$ the spin dispersion dives partialy
 into the
pair excitation continuum, so that excitations with
$k> 2q_{F}$ are strongly damped. For concentrations above the 
threshold $\delta^{*}\simeq 0.17$,
where the spin wave velocity equals the Fermi velocity, all the spin waves
lie in the pair excitation continuum, and therefore are completely damped.
This occurs at a concentration well below the value for which the spin
wave velocity would vanish.
One expects that long-range order will collapse at a concentration
$\delta_{c}<\delta^{*}$, implying a small value for the critical
 concentration, in
agreement with experimental data. The disappearence of the long-range 
magnetic order with doping will be discussed elsewhere$^{25}$. Aeppli \textsl{et al.}$^{5}$ and Hayden
\textsl{et al.}$^{6}$ investigated the spin dynamics of pure and doped
$La_{2-\delta}Sr_{\delta}CuO_{4}$, with $\delta=0.05$,
 and found that spin excitations
within the AF correlated regions in the doped material show softening and
damping with respect to the corresponding excitations in the pure material.
Aeppli \textsl{et} \textsl{al}. found that the spin wave velocity in the
doped material is renormalized
by a factor $0.74(\pm 0.08)$, while Hayden \textsl{et al}.
found a renormalization factor $0.60$. These values are to be compared
with the renormalization factor for the slope of the spin dispersion
around $q_{F}$, the momenta range associated to the AF 
correlated regions, therefore our result, $0.78$ for
 $\delta=0.05$, is in good agreement with experimental data.
 Experiments$^{7}$ at a much higher concentration, $\delta=0.14$,
 have also
 revealed a large
broadening of the spin excitations with doping.

For the transverse spin susceptibility we find a significant increase
with doping, having for $t/J=3$ a renormalization factor $Z_{\chi}=
${\large \ }$1.17$ for $\delta=0.02$, and $Z_{\chi}=${\large \ }$1.42$
for $\delta=0.05$. This effect is also due to the coherent motion of 
holes. When
long-range order is broken and the magnetic correlation length diverges, the
susceptibility of the system should be essentialy given by $\chi _{\perp }$.
An increase in the spin susceptibility with doping has in fact been observed
experimentally,$^{8-10}$ in agreement with our results. 
 Above the critical doping a gap gradually opens in the spin
excitation spectrum, and one may expect the magnetic correlation length in
that regime to be determined by the imaginary part of the spin dispersion.
According to our results this implies an inverse correlation length
proportional to $\mathrm{Im}\Pi^{+-} _{c,c}$ and therefore
$\xi \sim 1/\sqrt{\delta}$, precisely the scaling found experimentally.$^{24}$

Spin excitations of a weakly doped antiferromagnet have been investigated
elsewhere in the SCBA. In Ref.$21$, the present authors considered the
effect of the coherent motion of holes only, and showed that it generates
spin wave softening, in agreement with the results in the present work.
Other authors$^{17,20}$ claimed, however, that the coherent motion of holes
leads instead to stiffening of spin waves. This
contradicts our results, and may arise from approximations made in Refs.$17$
and $20$. The renormalization of the spin excitations has also been
calculated by another group,$^{23}$ but their calculation contains a
self-energy independent of $\delta$, a result difficult to understand.
 Becker and
Mushelknautz in Ref. $22$ also studied the effects of hole doping on the
spin excitations, but using a different technique. They found softening and
damping of the spin excitations, due to both the coherent and the incoherent
motion of holes, in agreement with our results.

In conclusion, we have shown that the magnetic properties of a
two-dimensional antiferromagnet are very sensitive to doping due to the
strong interaction between holes and spin waves, the coherent motion of 
holes leading to softening of the spin excitations, 
like the incoherent motion.

\bigskip
\noindent {\LARGE Appendix }

The contribution of the coherent motion of holes for the renormalization of
the spin waves excitations is given by $\mathrm{Re}\Pi _{c,c}^{+-}(\mathbf{k}
,\omega _{\mathbf{k}}^{0})$. From (5) and (6) one has

\begin{eqnarray*}
\mathrm{Re}\Pi _{c,c}^{+-}(\mathbf{k},\omega _{\mathbf{k}}^{0}) &=&a_{0}^{2}
\frac{1}{2N}\sum_{\mathbf{q}}\theta (|\mathbf{q}-\mathbf{k}
|-q_{F})\theta (q_{F}-|\mathbf{q}|)\times \\
&&\times \sum_{i=1}^{4}\left[ \frac{U^{+-}(\mathbf{k},\mathbf{k}-
\mathbf{q}-\mathbf{q}_{i})}{\omega _{\mathbf{k}}^{0}-(\varepsilon _{\mathbf{q
}-\mathbf{k}}-\varepsilon _{\mathbf{q}})}-\frac{U^{+-}(\mathbf{k},\mathbf{q}+
\mathbf{q}_{i})}{\omega _{\mathbf{k}}^{0}+(\varepsilon _{\mathbf{q}-\mathbf{k
}}-\varepsilon _{\mathbf{q}})}\right] ,
\end{eqnarray*}

\noindent where, $\varepsilon _{\mathbf{q}}=\mathbf{q}^{2}/2m$. Given that

\[
\sum_{i=1}^{4}U^{+-}(\mathbf{k},\mathbf{k}-\mathbf{q}+\mathbf{q}
_{i})=\sum_{i=1}^{4}U^{+-}(\mathbf{k},\mathbf{q}+\mathbf{q}
_{i})-(zt)^{2}\sum_{i=1}^{4}(\gamma _{\mathbf{q}-\mathbf{k}+\mathbf{q
}_{i}}^{2}-\gamma _{\mathbf{q}+\mathbf{q}_{i}}^{2}), 
\]

\noindent one has

\[
\mathrm{Re}\Pi _{c,c}^{+-}(\mathbf{k},\omega _{\mathbf{k}}^{0})=a_{0}^{2}\frac{
1}{N}\sum_{\mathbf{q}}\theta (|\mathbf{q}-\mathbf{k}
|-q_{F})\theta (q_{F}-|\mathbf{q}|)\left[ A(\mathbf{k},\mathbf{q})-B(\mathbf{
k},\mathbf{q})\right] , 
\]

\noindent where, 
\[
A(\mathbf{k},\mathbf{q})=\sum_{i=1}^{4}U^{+-}(\mathbf{k},\mathbf{q}+
\mathbf{q}_{i})\frac{2(\varepsilon _{\mathbf{q}-\mathbf{k}}-\varepsilon _{
\mathbf{q}})}{\left[ \left( \omega _{\mathbf{k}}^{0}\right) ^{2}-\left(
\varepsilon _{\mathbf{q}-\mathbf{k}}-\varepsilon _{\mathbf{q}}\right)
^{2}\right] } 
\]

\noindent and

\[
B(\mathbf{k},\mathbf{q})=(zt)^{2}\sum_{i=1}^{4}(\gamma _{\mathbf{q}-
\mathbf{k}+\mathbf{q}_{i}}^{2}-\gamma _{\mathbf{q}+\mathbf{q}_{i}}^{2})\frac{
1}{\left[ \omega _{\mathbf{k}}^{0}-(\varepsilon _{\mathbf{q}-\mathbf{k}
}-\varepsilon _{\mathbf{q}})\right] }. 
\]

For sufficiently small doping the spin wave velocity is larger than the
Fermi velocity, and therefore $A(\mathbf{k},\mathbf{q})\theta (|\mathbf{q}-
\mathbf{k}|-q_{F})\theta (q_{F}-|\mathbf{q}|)>0$. However, one has that $B(
\mathbf{k},\mathbf{q})>A(\mathbf{k},\mathbf{q})$, even in the
long-wavelength limit, $\mathbf{k}\ll 1$, where $B(\mathbf{k},\mathbf{q}
)\sim 2A(\mathbf{k},\mathbf{q})$. This implies that $\mathrm{Re}\Pi
_{c,c}^{+-}(\mathbf{k},\omega _{\mathbf{k}}^{0})<0$, and therefore softening
of the spin excitations, as given in (7). The inclusion of band anisotropy
does not qualitatively change our results.

\bigskip

\noindent {\LARGE References:}

\noindent $^{\*}$Electronic address: iveta@alf1.cii.fc.ul.pt

\noindent $^{1}$A. P. Kampf, Phys. Rep. \textbf{249}, 219 (1994).

\noindent $^{2}$F. C. Zhang and T. M. Rice, Phys. Rev. B \textbf{37}, 3759
(1988).

\noindent $^{3}$L. H. Tjeng, B. Sinkovic, N. B. Brookes, J. B. Goedkoop,
R. Hesper, E. Pellegrin, F. M. F. de Groot, S. Altieri, S. L. Hulbert,
E. Shekel, and G. A. Sawatzky, Phys. Rev. Lett. \textbf{78}, 1126 (1997).

\noindent $^{4}$R. J. Birgeneau and G. Shirane, in \textsl{Physical
Properties of High Temperature Superconductors}, edited by D. M. Ginzberg 
(World Scientific, Singapore, 1990).

\noindent $^{5}$G. Aeppli S. M. Hayden, H. A. Mook, Z. Fisk,
S.-W. Cheong, D. Rytz, J. P. Remeika, G. P. Espinosa, and A. S. Cooper,
Phys. Rev. Lett. \textbf{62}, 2052 (1989).

\noindent $^{6}$S. M. Hayden, G. Aeppli, H. Mook, D. Rytz, M. F. Hundley,
 and Z. Fisk, Phys. Rev. Lett. \textbf{66}, 821 (1991).

\noindent $^{7}$S. M. Hayden, G. Aeppli, H. A. Mook, T. G. Perring,
T. E. Mason, S.-W. Cheong, and Z. Fisk , Phys. Rev. Lett.
 \textbf{76}, 1344 (1996).

\noindent $^{8}$Y.-Q. Song, M. A. Kennard, K. R. Poppelmeier,
and W. P. Halperin , Phys. Rev. Lett. \textbf{70}, 3131 (1993).

\noindent $^{9}$T. Nakano, M. Oda, C. Manabe, N. Momono, Y. Miura,
 and M. Ido, Phys. Rev. B \textbf{49}, 16000 (1994).

\noindent $^{10}$S. Ohsugi, Y. Kitaoka and K. Asayama, Physica C
\textbf{282-287}, 1373 (1997).

\noindent $^{11}$S. Schmitt-Rink, C. M. Varma and A. E. Ruckenstein, Phys.
Rev. Lett. \textbf{60}, 2793 (1988).

\noindent $^{12}$C. L. Kane, P.A. Lee, and N. Read, Phys. Rev. B \textbf{39},
 6880 (1989).

\noindent $^{13}$G. Martinez and P. Horsch, Phys. Rev. B \textbf{44}, 317
(1991).

\noindent $^{14}$F. Marsiglio, A. E. Ruckenstein, S. Schmitt-Rink,
 and C. M. Varma, Phys. Rev. B \textbf{43}, 10882 (1991).

\noindent $^{15}$Z. Liu and E. Manousakis, Phys. Rev. B\textbf{45}, 2425
(1992).

\noindent \noindent $^{16}$E. Dagotto, Rev. Mod. Phys. B\textbf{66}, 763
(1994).

\noindent $^{17}$J. Igarashi and P. Fulde, Phys. Rev. B \textbf{45}, 12357
(1992). \strut

\noindent $^{18}$N. M. Plakida, V. S. Oudovenko and V. Yushanhai, Phys. Rev. 
\textbf{50}, 6431 (1994).

\noindent $^{19}$B. Kyung and S. Mukhin, Phys. Rev. B \textbf{55}, 3886
(1997).

\noindent $^{20}$G. Khaliullin and P. Horsch, Phys. Rev. B \textbf{47}, 463
(1993).

\noindent $^{21}$I.R. Pimentel and R. Orbach, Phys. Rev. B \textbf{46}, 2920
(1992).

\noindent $^{22}$K. W. Becker and M. Muschelknautz, Phys. Rev. B \textbf{48},
13826 (1993).

\noindent $^{23}$D. W. Murray and O. P. Suskov, Physica C \textbf{258}, 389
(1996).

\noindent $^{24}$R. J. Birgeneau, D. R. Gabbe, H. P. Jenssen, M. A. Kastner,
 P. J. Picone, T. R. Thurston, G. Shirane, Y. Endoh, M. Sato, K. Yamada,
  Y. Hidaka, M. Oda, Y. Enomoto, M. Suzuki, and T. Murakami,
  Phys. Rev. B \textbf{38}, 6614 (1988).

\noindent $^{25}$F. Carvalho Dias, I. R. Pimentel and R. Orbach
(unpublished).

\end{document}